\documentclass[amsmath]{revtex4}

\usepackage{graphicx}
\usepackage{subeqnarray}

\newcommand{\bra}[1]{\langle #1 | \,}
\newcommand{\ket}[1]{\, | #1 \rangle}
\newcommand{\de}{\delta}
\newcommand{\sgn}{\mathrm{sgn}}

\newcommand{\be}{\begin{equation}}
\newcommand{\ee}{\end{equation}}
\newcommand{\bea}{\begin{eqnarray}}
\newcommand{\eea}{\end{eqnarray}}
\newcommand{\besa}{\begin{subeqnarray}}
\newcommand{\eesa}{\end{subeqnarray}}
\newcommand{\bean}{\begin{eqnarray*}}
\newcommand{\eean}{\end{eqnarray*}}

\newcommand{\hb}{\hat{b}}
\newcommand{\hbd}{\hat{b}^{\dagger}}
\newcommand{\hn}{\hat{n}}
\newcommand{\hc}{\hat{c}}
\newcommand{\hcd}{\hat{c}^{\dagger}}

\newcommand{\Uc}{\mathcal{U}}

\begin{document}

\title{Exotic few-body bound states in a lattice}

\author{D. Petrosyan}
\author{M. Valiente}
\affiliation{Institute of Electronic Structure and Laser,
Foundation for Research and Technology--Hellas,
71110 Heraklion, Crete, Greece}

\begin{abstract}
Strongly-interacting ultra-cold atoms in tight-binding optical 
lattice potentials provide an ideal platform to realize the fundamental
Hubbard model. Here, after outlining the elementary single particle 
solution, we review and expand our recent work on complete 
characterization of the bound and scattering states of two
and three bosonic atoms in a one-dimensional optical lattice.
In the case of two atoms, there is a family of interaction-bound
``dimer'' states of co-localized particles that exists invariantly for
either attractive or repulsive on-site interaction, with the energy 
below or above the two-particle scattering continuum, respectively. 
Adding then the third particle---``monomer''---we find that, apart 
from the simple strongly-bound ``trimer'' corresponding to all three 
particles occupying the same lattice site, there are two peculiar 
families of weakly-bound trimers with energies below and above the
monomer--dimer scattering continuum, the corresponding binding 
mechanism being an effective particle exchange interaction.
\end{abstract}

\maketitle

\section{Introduction}
\label{sec:intro}

Among the tight-binding lattice models of condensed matter
physics\cite{SolStPh,Sadchev}, the Hubbard model\cite{Hubbard} plays 
a fundamental role. It describes particle tunneling between adjacent 
lattice sites as well as short range (contact) interaction between the 
particles on the same lattice site. Despite apparent simplicity, this 
model is very rich in significance and implications for the many
body physics on a lattice \cite{bosMI}. This is perhaps most profoundly 
manifested with numerous important experimental and theoretical 
achievements with cold neutral atoms trapped in deep optical 
lattice potentials \cite{optlattMIt,optlattMIe,OptLatRev}, wherein 
the Hubbard model is being realized with unprecedented accuracy. 

A remarkable Hubbard model phenomenon is the existence of 
stable repulsively-bound pairs of atoms in an optical lattice, 
as was experimentally demonstrated in Ref.~\cite{KWEtALPZ}. 
This seminal achievement has led to several theoretical studies
of the properties of interaction-bound atom pairs---``dimers''---in 
periodic potentials 
\cite{piilmolmer,pinymo,molmerdist,MVDP08jpb,MVDP08epl,MVDP09jpb,MV10,%
DPKLT,BSKLT,WHCh,JChS}. 

In a one-dimensional (1D) tight-binding lattice, two bosons can form 
a bound dimer\cite{KWEtALPZ,piilmolmer,pinymo,molmerdist,MVDP08jpb,MVDP08epl}
for any finite strength of the on-site interaction, be it an attraction 
or a repulsion. Next level of complexity corresponds to three bosons, 
which obviously can form a strongly-bound ``trimer'' with all 
three particles occupying the same lattice site. For large enough
on-site interaction strength, however, there are two more kinds
of weakly bound trimers \cite{MVDPAS10pra}. The corresponding binding 
mechanism turns out to be an effective particle exchange interaction 
between the dimer and the third particle--- ``monomer''---leading to  
symmetric and antisymmetric trimer states with energies slightly above 
and below the continuum of scattering states of (asymptotically) 
free dimer and monomer.

Below, after introducing the Bose-Hubbard model and outlining its 
elementary single particle solution, we review and expand our recent 
work \cite{MVDP08jpb,MVDPAS10pra} on the bound and scattering states 
of two and three bosonic atoms in a 1D optical lattice.

\section{The Model}
\label{sec:BHHam}

Cold bosonic particles in a 1D tight-binding lattice can be accurately 
described by the (second-quantized) Hubbard 
Hamiltonian \cite{Hubbard,bosMI,optlattMIt,optlattMIe,OptLatRev}
\be
H = - J \sum_{j} (\hbd_j \hb_{j+1} + \hbd_{j+1} \hb_j )
+ \frac{U}{2} \sum_{j} \hat{n}_j(\hat{n}_j-1) , \label{BHH}
\ee
where $\hbd_{j}$ ($\hb_{j}$) is the particles creation (annihilation)
operator and $\hn_j = \hbd_{j} \hb_{j}$ the number operator at $j$th
lattice site; $J$($>0$) is the inter-site tunneling, or hopping, rate; 
and $U$ is the on-site interaction, which can be attractive or repulsive.

\begin{figure}[ht]
\includegraphics[width=0.45\textwidth]{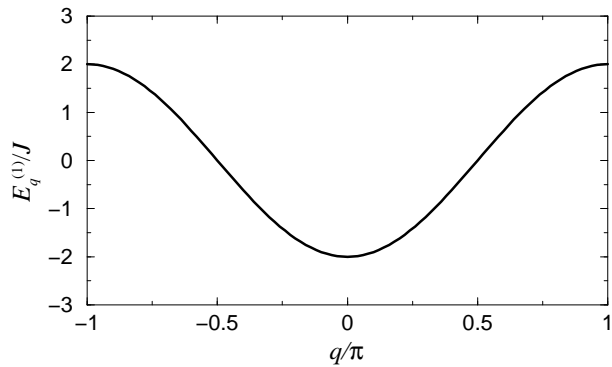}
\caption{Single particle Bloch band in a 1D tight-binding lattice.}
\label{fig:spBB}
\end{figure}

\paragraph{Single particle solution:} 
Denoting by $\ket{j}$ the state with a single
particle at the $j$th lattice site, the Hamiltonian 
reduces to 
\be
H^{(1)} = -J \sum_{j} (\ket{j}\bra{j+1} + \ket{j+1} \bra{j} ) . \label{HHamx}
\ee
Expanding the single-particle state vector as 
$\ket{\psi} = \sum_j \psi(j) \ket{j}$, the stationary Schr\"odinger 
equation $H^{(1)} \ket{\psi} = E^{(1)} \ket{\psi}$ 
leads to the difference equation
\be
-J\left[\psi(j+1)+\psi(j-1)\right] = E^{(1)} \psi(j),
\ee
which is satisfied by the (discrete) plane wave ansatz 
$\psi(j)=\psi_q(j)=\exp(i q j)$ for the wave function with
quasimomentum $q \in \Omega $ restricted to the first 
Brillouin zone $\Omega \equiv  [-\pi,\pi]$, while 
for the corresponding eigenenergy we obtain
$E_q^{(1)} = \epsilon(q) \equiv -2J\cos (q)$. 
The single-particle energy (Bloch) band, shown in Fig.~\ref{fig:spBB}, 
has, therefore, a width of $4J$.

Having reviewed the trivial single-particle solution \cite{SolStPh},
in the following Sections we present complete solutions of the much 
richer two- and three-body problems in a 1D lattice.

%%%%%%%%%%%%%%%%%%%%%%%TWO BODIES%%%%%%%%%%%%%%%%%%%%%%%%%%%%
\section{Two particles in a lattice}
\label{sec:2bs}

Although the Bose-Hubbard Hamiltonian (\ref{BHH}) corresponds to identical
bosons, which is our main concern here, it is nevertheless instructive 
to tackle the more general problem of two distinguishable particles
in a lattice \cite{molmerdist}, a simple limit of which produces 
the solution for two indistinguishable bosons. 

We thus consider two particles $A$ and $B$ having, in general, different 
hopping rates $J_A$ and $J_B$, respectively. The Hamiltonian describing 
their dynamics is the non-symmetrized (first-quantized) version of the 
Hubbard Hamiltonian (\ref{BHH}) acting on the two particle subspace: 
\be
H^{(2)} = 
-J_A \sum_{j_A} (\ket{j_A}\bra{j_A +1} + \ket{j_A+1} \bra{j_A }) 
-J_B \sum_{j_B} (\ket{j_B}\bra{j_B +1} + \ket{j_B +1}\bra{j_B }) 
+ U \sum_{j_A = j_B} \ket{j_A,j_B}\bra{j_A,j_B} , \label{H2}
\ee
where $j_A$ and $j_B$ denote the positions of particles $A$ and $B$, 
respectively. The eigenstates of $H^{(2)}$ can be expanded as
$\ket{\Psi}=\sum_{j_A,j_B} \Psi(j_A,j_B) \ket{j_A,j_B}$,
so that the Schr\"odinger equation $H^{(2)}\ket{\Psi}=E^{(2)}\ket{\Psi}$ 
leads to the difference equation
\be
-J_A \big[\Psi(j_A+1,j_B) + \Psi(j_A-1,j_B) \big] 
-J_B \big[\Psi(j_A,j_B+1) + \Psi(j_A,j_B-1) \big] 
+U \delta_{j_A,j_B}\Psi(j_A,j_B) = E^{(2)} \Psi(j_A,j_B) .\label{rec2body}
\ee
In order to solve the problem analytically, we need to transform the 
two-body difference equation (\ref{rec2body}) into a ``one-body'' problem. 
To that end, we define the center of mass $j_R=\frac{1}{2}(j_A + j_B)$ 
and relative $j_r = j_A - j_B$ coordinates and use for the two-particle
wave function the separation ansatz \cite{MVDPAS10pra,MV10}
\be
\Psi(j_A,j_B)= e^{i K j_R} e^{-i \beta_K j_r} \psi_K (j_r) ,\label{sepans}
\ee
where
\be
\tan(\beta_K) = \frac{J_A-J_B}{J_A+J_B} \tan(K/2) , \label{sepbeta}
\ee
with $K \in \Omega$ the center-of-mass quasimomentum. 
Note that for $J_A=J_B=J$ the separation ansatz (\ref{sepans}) reduces 
to that for identical particles \cite{MVDP08jpb,MVDP09jpb}. 
The resulting recursion relation now reads
\be
-J_K \big[\psi_K(j_r + 1) + \psi_K(j_r - 1) \big] + 
U \delta_{j_r,0} \psi_K(j_r) = E_K^{(2)} \psi_K(j_r) , \label{diffeqrel}
\ee
where $J_K = \sqrt{J_A^2 + J_B^2 + 2J_A J_B\cos(K)}$ is the collective 
hopping rate \cite{MV10,molmerdist}, which in the case of identical 
particles, $J_A=J_B=J$, reduces to the standard 
\cite{KWEtALPZ,piilmolmer,MVDP08jpb,MVDP09jpb} 
expression $J_K = 2J\cos(K/2)$.

Equation (\ref{diffeqrel}) admits two kinds of solutions, 
corresponding to the scattering states of asymptotically free particles 
and to the two-particle bound, or dimer, states.

\subsection{Scattering states}

Since the interaction between the particles is governed by a 
short-range---in this case a contact, $U \delta_{j_r,0}$---potential, 
its action amounts to a unitary phase shift (see below), while 
the spectrum of such solutions is given by the sum of the spectra 
for two free particles $A$ and $B$ with momenta 
$q_A = K/2 + k$ and $q_B = K/2 - k$:
\be
E_{K,k}^{(2)} = E_{q_A}^{(1)} + E_{q_B}^{(1)} = -2 J_K \cos(k) , \label{scatcont}
\ee
which spans the interval $E_{K,k}^{(2)} \in [-2J_K,2J_K]$. 
The corresponding symmetric scattering wave functions, for $\sin(k)\ne 0$, 
are given by
\be
\psi_{K,k}(j_r) = \cos(k |j_r| +\delta_{K,k}) , \label{scatwf}
\ee
which, upon substitution into Eq. (\ref{diffeqrel}),  
yields for the scattering phase shift $\delta_{K,k}$,
\be
\tan(\delta_{K,k}) = -\frac{U}{2J_K \sin (k)} . \label{tandelta0}
\ee
For $U\to 0$ we have non-interacting particles 
$\psi_{K,k}(j_r) = \cos (k |j_r|)$,
while for $U/J_K \to \pm \infty$ we obtain the fermionized solution
$\psi_{K,k}(j_r) = \sin (k |j_r|)$, whereby the two particles 
never occupy the same lattice site, $\psi_{K,k} (0) = 0$.

When $\sin (k) = 0$, the scattering wave functions (\ref{scatwf}) 
are no longer valid, and the lattice generalization of the continuum 
zero-energy solution of the Schr\"odinger equation apply \cite{pinymo,MV10}. 
At the bottom and the top of the scattering band, 
$E^{(2)}_{K,\mp } = \mp 2J_K$ ($k=0,\pi$), the corresponding solutions 
$\psi_{K,\mp}$ have the form
\besa
\label{scln}
\psi_{K,-}(j_r) &=& 1- \frac{|j_r|}{a_{K,-}} , \\
\psi_{K,+}(j_r) &=& (-1)^{j_r} \left( 1 - \frac{|j_r|}{a_{K,+}} \right) , 
\eesa
where $a_{K,\mp}$, with $a_{K,-}=-a_{K,+}$, are the scattering lengths, 
which are calculated by substituting Eqs. (\ref{scln}) into 
Eq. (\ref{diffeqrel}), resulting in $a_{K,-} = -2J_K/U$ (in units of
the lattice constant). Note that, since the scattering lengths 
are finite for any $U \neq 0$, there exists no ``zero-energy''
resonance in this model.\footnote{Contrast, however, with the 
extended Hubbard model in Refs.~\cite{MVDP09jpb,MV10}} 
Therefore, for every value of the total quasimomentum $K$ there is only 
one bound state, as we shall see below.

To illustrate the foregoing discussion, in Fig. \ref{fig:scatdims} we show 
the energy spectrum (\ref{scatcont}) for two identical bosons, $J_A=J_B=J$,
and the corresponding density of states defined via
\be
\rho(E,K) = \frac{L}{2 \pi} \frac{\partial k}{\partial E}
= \frac{L}{2 \pi} \frac{1}{\sqrt{[4 J \cos(K/2)]^2 - E^2}} , \label{rhoEK}
\ee
with $L$ a quantization length. As seen, the density of scattering states, 
i.e., the number of states per unit interval of energy, is 
lowest in the middle of the band, $E \simeq 0$ (and $K \simeq 0$), 
while for a given quasimomentum $K$ of the center of mass motion of the 
two particles, $\rho(E,K)$ increases rapidly as energy 
$E = E_{K,k}^{(2)}$ approaches its maximal and minimal values 
$\pm 2 J_K$.

\begin{figure}[ht]
\includegraphics[width=0.45\textwidth]{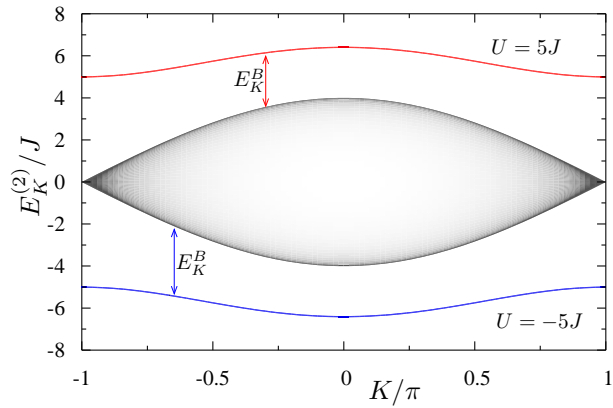}
\caption{Energies $E_{K}^{(2)}$ versus the center-of-mass 
quasimomentum $K$ for a pair of identical bosons in a 1D lattice. 
The continuum spectrum corresponds to energies~(\ref{scatcont})
of the scattering states, with the shading proportional to the 
density of states (\ref{rhoEK}).
The (blue) line below and the (red) line above the scattering 
band are, respectively, the energies (\ref{Ekadim}) of the 
attractively-bound dimer with $U = - 5J$ and 
the repulsively-bound dimer with $U = 5J$.}
\label{fig:scatdims}
\end{figure}

\subsection{Bound states}

The two-body bound states of Hamiltonian (\ref{H2}) are the solutions of 
Eq. (\ref{diffeqrel}) yielding normalizable relative coordinate wave function,
$\sum_{j_r} |\psi_K(j_r)|^2 < \infty$, with the corresponding energy below 
(for $U<0$) or above (for $U>0$) the scattering continuum at each value 
of the total quasimomentum $K$. Introducing into Eq. (\ref{diffeqrel})
the exponential ansatz $\psi_K(j_r) \propto \alpha_K^{|j_r|}$ yields the 
bound state energy $E_K^{(2)}=-J_K (\alpha_K + 1/\alpha_K)$, 
with $\alpha_K$ given by
\[ 
\alpha_K = \Uc_K -\sgn(U) \sqrt{\Uc_K^2 + 1} , \quad 
\Uc_K \equiv \frac{U}{2 J_K}.
\]
We see that $\alpha_K$ is real and also $|\alpha_K|<1$ for all $U \neq 0$. 
Hence, the relative coordinate wave function $\psi_K(j_r)$ is normalizable,
and the energy satisfies $|E_K^{(2)}| > 2J_K$ for all $K$. 
Explicitly, the energy and the normalized wave function
for the bound dimer are given by
\bea
E_K^{(2)} &=& \sgn(U) \sqrt{U^2 + 4 J_K^2}, \label{Ekadim} \\
\psi_K (j_r) &=&  \frac{\sqrt{|\Uc_K|}}{\sqrt[4]{\Uc_K^2 + 1}} \,
\Big( \Uc_K -\sgn(U)\sqrt{\Uc_K^2 + 1}  \Big)^{|j_r|}. \label{psiadim}
\eea

In Fig. \ref{fig:scatdims} we show the energies $E_{K}^{(2)}$  
for a pair of identical bosons, $J_A = J_B = J$, interacting via 
on-site attractive, $U <0$, or repulsive, $U > 0$, potential. 
The binding energies $E_K^{B}$ are defined with respect to the 
edges of the scattering band (\ref{scatcont}). Note that in 
the case of repulsive interaction, $U > 0$ ($\alpha_K < 0$), 
the sign of the wave function (\ref{psiadim}) alternates between 
the neighboring sites $j_r$. Remarkably, when $|K| = \pi$, and 
thereby $J_K = 0$, the relative coordinate wave function $\psi_K (j_r)$ 
is completely localized at $j_r=0$ for any $U \neq 0$.

Clearly, for a given value of the dimer quasimomentum $K$, 
the stronger is the on-site interaction $|U|$, the smaller is 
the extent of the wave function $\psi_K (j_r)$, meaning that the 
constituent particles are stronger co-localized. For $|U| \gg J$, 
the dimer energy in Eq. (\ref{Ekadim}) can be approximated as
\be
E_K^{(2)} \simeq \mathcal{E}^{(2)} - 2 J^{(2)}  \cos(K) , \label{Ekdim}
\ee
where the first term $\mathcal{E}^{(2)} \equiv [U - 2 J^{(2)}]$ represents 
the dimer ``internal energy'', while the second term 
$\epsilon^{(2)}(K) \equiv - 2 J^{(2)} \cos(K)$ is the kinetic energy 
of a dimer with quasimomentum $K$ and an effective tunnelling rate 
$J^{(2)} \equiv - 2 J^2 /U$.

\section{Three particles in a lattice}

Building on the solution of the two-body problem, in this section we 
consider three bosonic atoms is an optical lattice described by  
Hamiltonian~(\ref{BHH}).

\subsection{Spectrum of scattering states}

With the expertise gained from the previous section, we can readily 
deduce that in the case of three particles there are two distinct 
scattering continua (see Fig. \ref{fig:3bspct} left panel). 
The first is the three-body scattering continuum of three (asymptotically)
free particles, with the energy given by the sum of single-particle 
bands, $E_{c3} = \epsilon(k_1) + \epsilon(k_2)+ \epsilon(K-k_1-k_2)
\equiv \epsilon(k_1,k_2,K-k_1-k_2)$, where each particle quasimomentum 
$k_j \in \Omega$ is in the first Brillouin zone $\Omega \equiv  [-\pi,\pi]$ 
and $K=k_1+k_2+k_3 \pmod {2\pi} $ is the total quasimomentum. 
The second is the two-body scattering continuum of a bound pair (dimer) 
and a free particle (monomer), with energy 
$E_{c2} =\sgn(U)\sqrt{U^2+ [4 J \cos(Q/2)]^2} - 2J \cos(K-Q)$, 
where the first term is the energy of a dimer, Eq.~(\ref{Ekadim}), 
with quasimomentum $Q$.

\subsection{Bound states}

We seek the bound states $\ket{\Psi}$ of three bosons 
in momentum representation,
\be
\ket{\Psi} = \frac{1}{(2\pi)^{3/2}} \iiint_{\Omega^3} \!\! dk_1 dk_2 dk_3 \,
\Psi(k_1,k_2,k_3) \ket{k_1,k_2,k_3},
\ee
where the wave function $\Psi(k_1,k_2,k_3)$ is symmetric with respect to 
exchange of any pair of particles. From the stationary Schr\"odinger equation 
$H\ket{\Psi}=E\ket{\Psi}$, using the conservation of total quasimomentum $K$, 
we obtain \cite{mattis}  
\be
\Psi(k_1,k_2,k_3) = -\frac{M(k_1)+M(k_2)+M(k_3)}
{\epsilon(k_1,k_2,k_3)-E},
\ee
where functions $M(k)$ satisfy the 1D Mattis integral equation \cite{mattis}
\be
M(k) [1+I_E(k)] 
= -\frac{U}{\pi}\int_{-\pi}^{\pi} \!\! dq \, 
\frac{M(q)}{\epsilon(k,q,K-k-q) - E }, \label{Minteq}
\ee
with $I_E(k)$ being a generalized Watson integral \cite{watson}
\[
I_E(k) \equiv  \frac{U}{2\pi} \int_{-\pi}^{\pi}
\!\! dq \, \frac{1}{\epsilon(k,q,K-k-q)-E} 
= -\frac{\sgn [ E -\epsilon(k) ] U}
{\sqrt{ [ E-\epsilon(k) ]^2-16J^2\cos^2[(K-k)/2]}}.
\]
Equation (\ref{Minteq}) can be cast as a homogeneous Fredholm 
equation of the second kind with eigenvalue $\lambda=1$. 
Hence, for a given $U/J$ and fixed $K$, it is a nonlinear 
equation for energy $E$, which we solve numerically. 

\begin{figure}[b]
\includegraphics[width=0.7\textwidth]{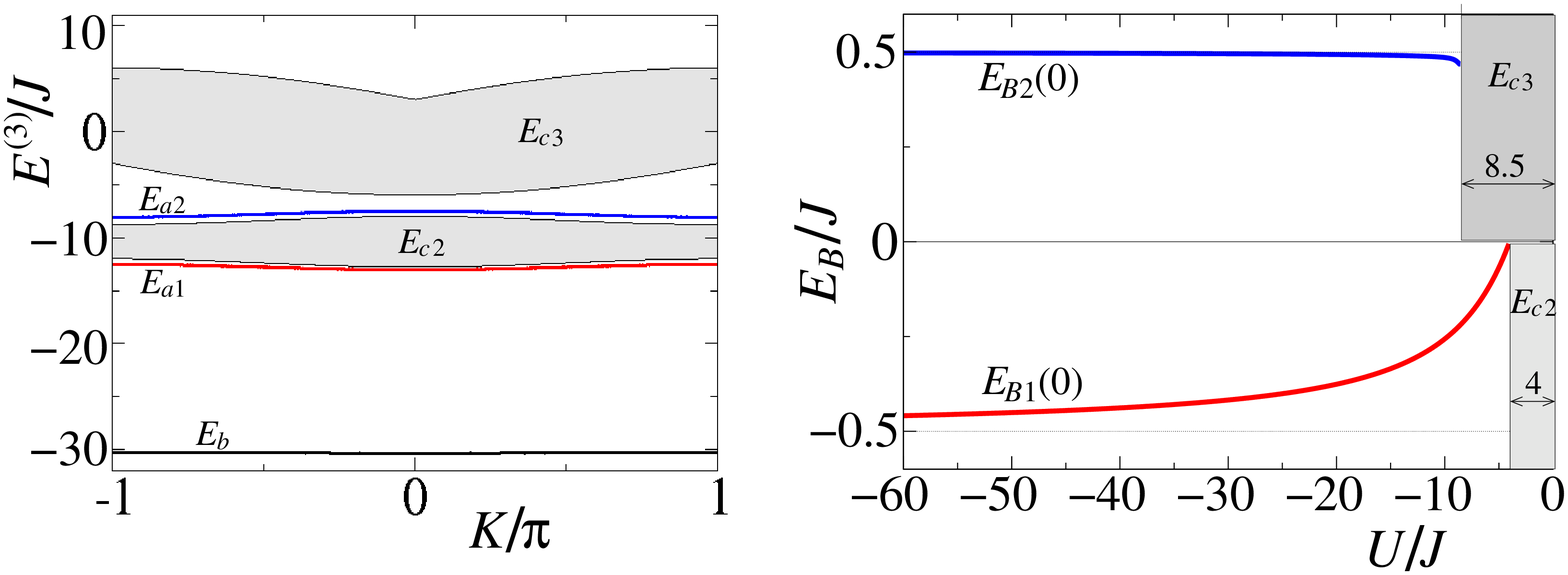}
\caption{Left: Full three-particle energy spectrum of Hamiltonian (\ref{BHH}) 
with $U=-10J$, versus the total quasimomentum $K$. All bound states are 
obtained via exact numerical solution of Eq.~(\ref{Minteq}).
Right: Binding energies $E_B$ for the off-site (weakly-bound) 
trimers at $K=0$ versus the interaction strength $U <0$.}
\label{fig:3bspct}
\end{figure}

The full three-body spectrum of Hamiltonian (\ref{BHH}) 
is shown in Fig.~\ref{fig:3bspct} left panel, with the bound state 
energies denoted by $E_b$, $E_{a1}$ and $E_{a2}$. For concreteness, 
here we consider attractive interaction, $U<0$, but note 
that our results equally apply to the case of repulsive 
interaction, $U>0$ \cite{MVDPAS10pra,MV10}.

In complete analogy with the two-body problem, the Bose-Hubbard 
Hamiltonian (\ref{BHH}) with $|U|/J\gg 1$ has a very narrow band 
of on-site bound states, corresponding to three tightly 
bound bosons co-localized on the same lattice site, with energies 
$E_b \approx 3U$ far from both scattering continua \cite{mattis}. 
But as also seen in Fig.~\ref{fig:3bspct}, in a 1D lattice, bosons can 
form two new kinds of three-body bound states whose energies $E_{a1}$ 
and $E_{a2}$ lie below and above the two-body continuum $E_{c2}$.
Some properties of these states can be deduced by energy considerations. 
First, these are not on-site bound states, 
since their energies $E_{a1(2)} \simeq U + O(J)$ are far from $3U$. 
Next, their binding energies, with respect to the $E_{c2}$ band, 
are $E_{B1(2)} \lesssim \mp J/2$, which suggests that these are 
off-site weakly-bound states of a dimer and a monomer. Note that 
the state above the two-body continuum is bound stronger than the 
state below the continuum. Finally, they are not Efimov states which 
can exist only in 3D systems near two-boson resonances \cite{mattis,Efimov}.

As can be seen from Fig.~\ref{fig:3bspct} right panel, where we plot 
the binding energies $E_{B1(2)}$ at quasimomentum $K=0$, there are 
thresholds for the existence of full bands ($K \in [-\pi, \pi]$) 
of the off-site bound states.  
For the trimer below the two-body continuum, the binding energy 
vanishes when $|U|\approx 4 J$: at this critical value of $U$
the trimer energy $E_{a1}$ approaches the edge of the dimer--monomer
scattering continuum $E_{c2} = - \sqrt{U^2 +16 J^2} -2 J$. 
On the other hand, the $K=0$ trimer above the two-body continuum 
ceases to exist already for $|U| \approx 8.5 J$, since then 
its energy $E_{a2}$ approaches the bottom of the three-body 
continuum $E_{c3} = 3 \epsilon(0) = - 6J$ 
(the two continua, $E_{c2}$ and $E_{c3}$, overlap for $|U| \leq 8J$). 
Thus, at $K=0$, the trimer state with energy $E_{a2}$ starts to 
appear well in the strong interaction regime, while for larger $K$ 
the threshold is smaller: $|U| \approx 4 J$ for $|K| \to \pi$.

\subsection{Effective model}

Since for strong on-site interaction $|U|/J \gg 1$ the dimer is essentially
unbreakable, the off-site trimers with energies close to $U$ must be 
bound due to a mechanism different from the on-site interaction alone. 
To identify such a mechanism, we derive an effective perturbative model,
valid for $|U|/J>8$, describing two distinguishable, hard-core 
particles---the dimer and the monomer. To second order in the 
tunneling rate $J$, the effective Hamiltonian reads
\be
H_{\mathrm{eff}} = H_1 + H_2 + H_{\mathrm{int}} , \label{effHam}
\ee
where
\[
H_1 = - J \sum_{j} (\hbd_j \hb_{j+1} + \mathrm{H.c.})
\] 
describes the single monomer;
\[
H_2 = \mathcal{E}^{(2)} \sum_{j} \hat{m}_j -
J^{(2)} \sum_{j} (\hcd_j \hc_{j+1} + \mathrm{H.c.})
\]
is the Hamiltonian for a dimer ({\it cf.} Eq.~(\ref{Ekdim})), 
with $\hcd_j$ ($\hc_{j}$) being the dimer creation (annihilation) 
operator and $\hat{m}_j = \hcd_j \hc_j$ the number operator at site $j$;
and finally 
\[
H_{\mathrm{int}} =  V^{(2)} \sum_j \hat{m}_j\hat{n}_{j \pm 1}
- W \sum_j (\hcd_{j+1} \hc_j \hbd_j \hb_{j+1} + \mathrm{H.c.})
\]
describes effective interactions between the dimer and the monomer, 
including a weak nearest-neighbor interaction $V^{(2)} = -7 J^2/2U$, 
and an exchange interaction with the rate $W = 2 J$. As we will see 
below, it is the exchange term that is responsible for the formation 
of the off-site trimers.

\begin{figure}[b]
\includegraphics[width=0.6\textwidth]{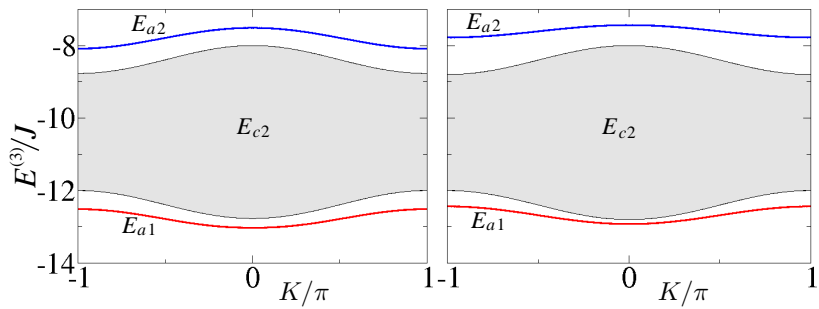}
\caption{Left: Magnified part of the exact three-particle spectrum 
of Fig.~\ref{fig:3bspct}. 
Right: Dimer--monomer spectrum of the effective Hamiltonian (\ref{effHam}) 
with the two bound states obtained via numerical solution of Eq.~(\ref{dmPsi}).}
\label{fig:3bexef}
\end{figure}

In Fig.~\ref{fig:3bexef} right panel, we plot the spectrum of the 
effective Hamiltonian~(\ref{effHam}), which contains two bound states 
with energies $E_{a1}$ and $E_{a2}$ below and above the two-body scattering
continuum $E_{c2} = \mathcal{E}^{(2)} + \epsilon^{(2)}(Q) + \epsilon(K-Q)$.
These dimer--monomer bound states are obtained using the Schr\"odinger 
equation for the two-body wave function $\Psi(Q,k)$ in momentum space, 
which leads to the integral equation
\be
\Psi(Q,k)= -\frac{1}{2\pi} \int_{-\pi}^{\pi} \!\! dq \,
\frac{U_{12} + V_{\mathrm{cos}}(Q,q) + V_{\mathrm{sin}}(Q,q)}
{\mathcal{E}^{(2)} + \epsilon^{(2)}(q) + \epsilon(K-q) - E}
\Psi(q,k), \label{dmPsi}
\ee
where $K = Q+k $ is the total quasimomentum,
$V_{\mathrm{cos}}(Q,q) = [2V^{(2)}\cos(q)- 4J \cos(K-q)]\cos(Q)$ and
$V_{\mathrm{sin}}(Q,q) \equiv V_{\mathrm{cos}}(Q,q)$ with $\cos \to \sin$,
while $U_{12} (\to \infty)$ is an artificial dimer--monomer on-site
interaction imposing the hard-core condition on Hamiltonian (\ref{effHam}).
Equation~(\ref{dmPsi}) reduces to a non-linear equation for the 
energy $E$ solving which (numerically) we obtain $E_{a1}$ and $E_{a2}$. 
Comparison with the exact spectrum on the left panel of Fig.~\ref{fig:3bexef} 
reveals good agreement: the continuum spectra are indistinguishable, 
while the small but noticeable differences in the bound-state energies 
are associated with the internal structure of the dimer, not 
accounted for by the effective model, and they gradually 
disappear with increasing the on-site interaction strength $U$.

\subsubsection{Analytic solutions}

There are two important cases, corresponding to the maximal ($K=\pi$) 
and minimal ($K=0$) total quasimomentum $K$, for which the bound and 
scattering states of the effective Hamiltonian (\ref{effHam}) can be 
calculated analytically employing the method of Sec. \ref{sec:2bs}. 
To that end, we expand the two-particle eigenstates in coordinate 
basis $\ket{\Psi}=\sum_{j_1 \neq j_2} \Psi(j_1,j_2) \ket{j_1,j_2}$
with the wave function in the form 
$\Psi(j_1,j_2)= e^{iK j_R} e^{-i \beta_K j_r} \psi_K (j_r)$, 
where $j_R \equiv \frac{1}{2} (j_1 + j_2)$, $j_r \equiv j_1 - j_2$, 
with $j_1$ and $j_2$ being the lattice positions of the monomer 
and dimer, and $\tan(\beta_K) = \tan{(K/2)} [J-J^{(2)}]/[J+J^{(2)}]$.   
For the relative coordinate wave function $\psi_K (j_r)$, 
imposing the hard-core condition $\psi_K(0)=0$, we then 
obtain the difference relations
\bea
J_K \psi_K(\pm 2) + W_K \psi_K(\mp 1) + [\bar{E} - V^{(2)}] \psi_K(\pm 1) &=& 0 ,
\nonumber \\
J_K [\psi_K(j_r+1)+\psi_K(j_r-1) ] + \bar{E} \psi_K(j_r ) &=& 0 , \label{phieqs}
\eea
with $|j_r| > 1$, $J_K \equiv \sqrt{J^2 + J^{(2)2} + 2 J J^{(2)} \cos (K)}$,
$W_K \equiv W \cos (K)$, and $\bar{E} \equiv E - \mathcal{E}^{(2)}$.

\paragraph{Scattering solutions:}

Using the standard ansatz 
$\psi_{K,k}^{(+)}(j_r) = \cos \big( k|j_r| + \de_{K,k}^{(+)} \big)$ and 
$\psi_{K,k}^{(-)}(j_r) = \sgn(j_r) \cos \big( k|j_r| + \de_{K,k}^{(-)} \big)$ 
with $k$ the relative quasimomentum, for the corresponding phase shifts 
$\de_{K,k}^{(\pm)}$ of the symmetric ($+$) and antisymmetric ($-$) 
scattering wave functions we obtain 
\be
\tan(\delta_{K,k}^{(\pm)})= 
\frac{J_K \cos(2k) + [\bar{E} \pm W_K - V^{(2)}] \cos(k)}
{J_K \sin(2k)+[\bar{E} \pm W_K - V^{(2)}]\sin(k)} , \label{tandelta} 
\ee
with $\bar{E} = E_{c2} - \mathcal{E}^{(2)} = -2 J_K \cos(k)$. 
Note that in the limit of $|U| \to \infty$, as the the nearest neighbour 
interaction $V^{(2)}$ and the dimer hopping $J^{(2)}$ tend to zero 
($J_K \to J$), Eq. (\ref{tandelta}) holds for all $K = k$.
The full scattering wave function is given by a superposition 
$\psi_{K,k}(j_r) = A \, \psi_{K,k}^{(-)}(j_r) + B \, \psi_{K,k}^{(+)}(j_r)$,
which, upon expressing through incident, reflected and transmitted waves,
\[
\psi_{K,k}(j_r ) = 
\begin{cases}
  e^{i k j_r} + r_K \, e^{-i k j_r}  \quad  & (j_r < 0)  \\
  t_K \, e^{i k j_r} & (j_r > 0)  
\end{cases} ,
\]
leads to $A/B = -e^{-i [ \de_{K,k}^{(+)} - \de_{K,k}^{(-)} ]}$. 
For the reflection $r_K(k)$ and transmission $t_K(k)$ amplitudes we then obtain 
$r_K,t_K = \frac{1}{2} \big[ e^{2 i \de_{K,k}^{(+)}} \pm  e^{2 i \de_{K,k}^{(-)}} \big]$,
and the transmission and reflection probabilities are given by 
$T_K =|t_K|^2 = \sin^2 \big( \de_{K,k}^{(+)} - \de_{K,k}^{(-)} \big)$
and $R_K =|r_K|^2 = \cos^2 \big( \de_{K,k}^{(+)} - \de_{K,k}^{(-)} \big)$. 

\begin{figure}[ht]
\includegraphics[width=0.45\textwidth]{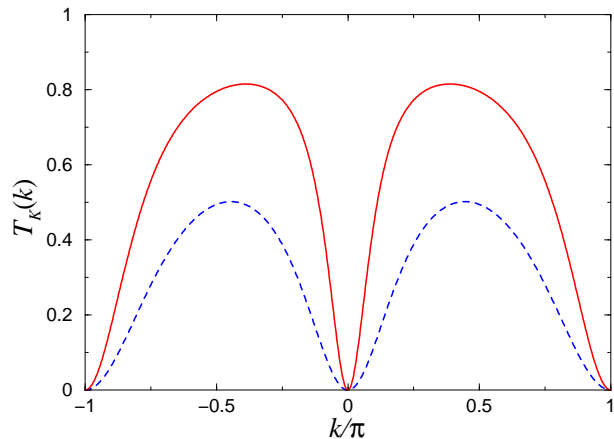}
\caption{Transmission probability $T_K(k)$ of a single particle with 
relative quasimomentum $k$ through a bound dimer, for $U/J=-10$ and
total quasimomenta $K=0$ (red solid line) and $|K|=\pi$ (blue dashed line).}
\label{fig:Trns}
\end{figure}

In Fig.~\ref{fig:Trns} we plot $T_K(k)$ for $U/J=-10$ at total 
quasimomenta $K=0$ and $|K|= \pi$. The transmission spectra for the 
intermediate values of $K$ lies in between the curves 
for $K=0$ and $K = \pi$. We observe the maximum transmission in the 
vicinity of $k = \pm \pi/2$, where $T_K(\pi/2)$ ranges from 50\% to 80\%.
With increasing the interaction strength $U$, we find that 
the maximum transmission saturates at around 64\% for all values of $K$,
which should be contrasted with the results of Ref. \cite{JChS}.

%%%%%%%%%%%%%%%%%
%%%%%%%%%%%%%%%%%

\paragraph{Bound solutions:}

The dimer--monomer bound states are obtained 
from Eq.~(\ref{phieqs}) using the exponential ansatz
$\psi_K(j_r>0) \propto \alpha_K^{j_r-1}$ and $\psi_K(-j_r) = \pm \psi_K(j_r)$,
which yields 
\be\alpha_K^{(\pm)} = - \frac{J_K}{V^{(2)} \mp W_K} 
\ee
for the symmetric ($+$) and antisymmetric ($-$) wave function
of the bound state ($|\alpha_K| < 1$), with the corresponding energy
$\bar{E}_{a1(2)} = - J_K [1+ (\alpha_K^{(\pm)})^2]/\alpha_K^{(\pm)}$.

It is now easy to see that without the exchange interaction there 
would be no dimer--monomer bound states (for any $K$). 
Indeed, this hypothetic ($W =0$) problem is exactly solvable for all $K$, 
and for two hard-core bosons with nearest-neighbour interaction $V^{(2)}$ 
there could be only one bound state\cite{MVDP09jpb} when 
$|\alpha_K| = |J_K/ V^{(2)}| < 1$, which cannot be satisfied in 
the range of validity ($|U|/J>8$) of the effective model, Eq. (\ref{effHam}).
The effective nearest-neighbour interaction is, however, responsible 
for the asymmetry in the binding energies 
$E_{B1(2)} = \bar{E}_{a1(2)} \mp 2 J_K$ of the exchange-bound trimers 
below and above the continuum $E_{c2}$ (see Fig. \ref{fig:3bspct} right panel).
With increasing the on-site interaction $U$, the binding energies attain 
the universal limits $E_{B1(2)} \to \mp J/2$ which remain valid for all $K$,
since $J^{(2)} \to 0$ as $|U/J|\to \infty$. We thus conclude that
the effective particle exchange interaction binds the dimer and the
monomer into the weakly bound trimer.

\section{Conclusions}

In this paper, we have presented complete solutions for the 
one-, two- and three-body problems in a 1D tight-binding lattice 
described by the Bose-Hubbard model. 

For the case of two bosons, we have derived the scattering states and 
the bound states of co-localized particles, which we termed as dimers.
The corresponding binding mechanism is provided by the on-site 
interaction between the particles. Perhaps counterintuitively, 
these bound dimer states exist not only for attractive, but also for 
repulsive interactions \cite{KWEtALPZ}. In free space, the repulsive 
interaction would inevitably lead to the pair separation, or dissociation, 
whereby the potential energy of the repulsion is converted into 
the kinetic energy of the free particles. In the lattice, however, 
the kinetic energy of each particle cannot take on arbitrary values,
but is restricted to the values in the allowed Bloch band, which 
itself is bounded both from below and from above. Therefore, in the 
absence of energy dissipation, a pair of co-localized particles 
interacting even via repulsive potential is destined to stay bound 
together as a dimer, simply because there are no available free-particle
energy states to which the dimer can dissociate.  

For three bosons, we have found three families of 
trimers---bound states of three particles. The first strongly-bound 
trimer, being an analog of the dimer, corresponds to all 
three particles occupying the same lattice site and bound by the
on-site interaction. The other two families of trimers are 
weakly-bound with energies just below and above the two-body 
scattering continuum of a single particle (monomer) and an 
interaction-bound dimer. Intuitively, these trimer states correspond 
to symmetric and antisymmetric states of a dimer and monomer at the 
neighboring lattice sites interacting with each other via an effective 
(particle) exchange interaction, which is responsible for their binding.

The phenomena discussed above are pertinent to the experiments
with cold bosonic atoms in optical lattices \cite{OptLatRev}.
Studying larger number of bosons in a lattice might reveal other 
exotic bound states, while longer range interactions, such as those
between dipolar atoms \cite{DipAt}, or molecules \cite{DipMol},
will certainly play an important role in the formation of few-body 
bound states \cite{MVDP09jpb,MV10,MVthes}.

\section*{Acknowledgements}

This work was supported by the EC Marie-Curie Research Training Network EMALI.

\end{document}